\documentclass[conference,letterpaper]{IEEEtran}

\usepackage[utf8]{inputenc} 
\usepackage[T1]{fontenc}
\usepackage{url}
\usepackage{ifthen}
\usepackage{cite}
\usepackage[cmex10]{amsmath} % Use the [cmex10] option to ensure complicance
\usepackage{graphicx}
\usepackage{hyperref}
\usepackage{amsmath}
\usepackage{amssymb}
\usepackage{bm}
\usepackage{bbm}
\usepackage{algorithmicx}
\usepackage{algorithm}
\usepackage{algpseudocode}
\usepackage{array}
\usepackage{booktabs}
\usepackage{multirow}
\usepackage{makecell}
\usepackage{mathtools}
\usepackage{amsthm}

\newtheorem{definition}{Definition}
\newtheorem{theorem}{Theorem}

\newcommand{\A}{\mathcal{A}}
\newcommand{\I}{\mathcal{I}}

\newcommand{\List}{\mathcal{L}}

\DeclareMathOperator*{\TBP}{TBP}
\DeclareMathOperator*{\IEE}{IEE}
\DeclareMathOperator*{\CRC}{CRC}
\DeclareMathOperator*{\Candidate}{Candi}

\DeclareMathOperator{\Free}{free}
\begin{document}

\title{An Efficient Algorithm for Designing Optimal CRCs for Tail-Biting Convolutional Codes \thanks{Thanks NSF XXXXXxXXXX}}

% %%% Single author, or several authors with same affiliation:
% \author{%
%   \IEEEauthorblockN{Stefan M.~Moser}
%   \IEEEauthorblockA{ETH Zürich\\
%                     ISI (D-ITET)\\
%                     CH-8092 Zürich, Switzerland\\
%                     Email: moser@isi.ee.ethz.ch}
% }

%%% Several authors with up to three affiliations:
\author{\IEEEauthorblockN{Hengjie~Yang, Linfang~Wang, Vincent Lau and Richard~D.~Wesel}
\IEEEauthorblockA{
Department of Electrical and Computer Engineering\\
University of California, Los Angeles, Los Angeles, CA 90095, USA\\
Email: \{hengjie.yang, lfwang, vincentlau, wesel\}@ucla.edu}
}

\maketitle

%%%%%%
%% Abstract: 
%% If your paper is eligible for the student paper award, please add
%% the comment "THIS PAPER IS ELIGIBLE FOR THE STUDENT PAPER
%% AWARD." as a first line in the abstract. 
%% For the final version of the accepted paper, please do not forget
%% to remove this comment!
%%
\begin{abstract}
Cyclic redundancy check (CRC) codes combined with convolutional codes yield a powerful concatenated code that can be efficiently decoded using list decoding.  To help design such systems, this paper presents an efficient algorithm for identifying the distance-spectrum-optimal (DSO) CRC polynomial for a given tail-biting convolutional code (TBCC) when the target undetected error rate (UER) is small. Lou \emph{et al.} found that the DSO CRC design for a given zero-terminated convolutional code under low UER is equivalent to maximizing the  undetected minimum distance (the minimum distance of the concatenated code). This paper applies the same principle to design the DSO CRC for a given TBCC under low target UER. Our algorithm is based on partitioning the tail-biting trellis into several disjoint sets of tail-biting paths that are closed under cyclic shifts. This paper shows that the tail-biting path in each set can be constructed by concatenating the irreducible error events (IEEs) and circularly shifting the resultant path. This motivates an efficient collection algorithm that aims at gathering IEEs, and a search algorithm that reconstructs the full list of error events with bounded distance of interest, which can be used to find the DSO CRC. Simulation results show that DSO CRCs can significantly outperform suboptimal CRCs in the low UER regime.
\end{abstract}

%% The paper must be self-contained. However, if you are referring to
%% a full version for checking certain proofs, please provide the
%% publically accessible location below.  If the paper is completely
%% self-contained, you can remove the following line from your
%% submission.

{\let\thefootnote\relax\footnote{{This research is supported by National Science Foundation (NSF) grant CCF-2008918 and Physical Optics Corporation (POC). Any opinions, findings, and conclusions or recommendations expressed in this material are those of the author(s) and do not necessarily reflect views of the NSF or POC.}}}

\section{Introduction}
\label{sec: introduction}

Tail-biting convolutional codes (TBCCs) are simple and powerful codes in the short blocklength regime. Unlike the conventional zero-terminated convolutional code (ZTCC) whose trellis paths all begin and end in the zero state, a TBCC only requires that each trellis path starts and ends at the same state.  This avoids the need for termination bits.  TBCCs were first proposed by Ma and Wolf \cite{Ma1986} as a modified version of the ZTCC to eliminate the rate loss caused by termination bits. Solomon and Tilborg \cite{Solomon1979} demonstrated the intriguing relation that any TBCC can be transformed into a quasi-cyclic code and conversely, many quasi-cyclic codes can be viewed as a TBCC with a small constraint length. Subsequently, it was shown that any linear block code can correspond to a tail-biting (TB) trellis representation and the code represented by such trellis is called a TB code \cite{Calderbank1999,Koetter2002}. The significance of TB codes lies in the fact that they achieve the best minimum distance of codes in the short-to-medium blocklength regime \cite{Ma1986, Stahl1999,Bocharova2002}.

Since the advent of TBCCs and TB codes, several authors  proposed a variety of algorithms to decode a TBCC or a TB code, e.g., \cite{Ma1986,Wang1989,Cox1994,Anderson1998,Shao2003,Chen2008,Williamson2014}. These algorithms are based on either maximum likelihood (ML) or maximum a posteriori (MAP) criteria. For ML decoding algorithms, the wrap-around Viterbi algorithm (WAVA) \cite{Shao2003} achieves the near-ML performance with the minimum complexity.

Cyclic redundancy checks (CRCs) are commonly used to detect whether a codeword is correctly received. Recently, with the development of 5G, CRC-aided list decoding of finite blocklength codes has received increasing popularity.  CRC-aided list decoding can significantly help improve the code performance, e.g., \cite{Niu2012,Yang2018,Coskun2019, Liang2019}. Lou \emph{et al.} \cite{Lou2015} first designed the optimal CRC for a given ZTCC such that the concatenated CRC-ZTCC achieves the minimum undetected error rate (UER) when the target UER is low. The CRC they designed can be referred to as the \emph{distance-spectrum-optimal} (DSO) CRC in the sense that the upper bound of the UER  characterized by the full undetected distance spectrum is minimized and the upper bound is close to the true UER when the target UER is set low. However, DSO CRC design for a given TBCC is still missing from the literature. It is remarkable that a simple suboptimal CRC design \cite{Liang2019} can nearly achieve the random coding union bound of Polyanskiy \emph{et al.} \cite{Polyanskiy2010}.

The DSO CRC design principle under a low target UER parallels that of Lou \emph{et al.}, which is equivalent to maximizing the \emph{undetectable minimum distance} of the overall concatenated code. To this end, the first step is to gather a sufficient number of error events, i.e., TB paths, of distances less than some threshold. This can be accomplished by the \emph{collection algorithm}. Then, the \emph{search algorithm} is employed to find the DSO CRC polynomial that maximizes the undetectable minimum distance. For TBCCs, a trivial collection algorithm is to perform Viterbi search separately at each possible initial state to find all error events of a bounded distance. However, such an algorithm will be \emph{inefficient} in collecting TB paths for a family of objective trellis lengths. If the objective trellis length changes to a smaller value, one has to redo the above procedure from scratch.

% First, the buffer size of caching intermediate paths will increase exponentially as the depth increases. Second, a significant portion of those intermediate paths will be discarded eventually due to the distance constraint, causing an inefficient use of memory. Third, the TB paths found by the trivial collection algorithm cannot be adapted to other trellis lengths.

Unlike the trivial algorithm, this paper provides an \emph{efficient} algorithm that supports the DSO CRC design of a given TBCC for a family of objective trellis lengths. The algorithm is based on partitioning the TB trellis into several disjoint sets of TB paths that are closed under cyclic shifts. Specifically, for a feedforward convolutional encoder with $v$ memory elements and a specified blocklength, its TB trellis can be described as the union of all TB paths of the required length that start and end at any of the $2^v$ states. Each TB path can be categorized by a state through which it traverses. Let $\TBP(0)$ be the set of TB paths that traverse through state $0$. Then, recursively define $\TBP(i)$, $1\le i\le 2^v-1$, as the set of TB paths that traverse through state $i$ but not through $0,1,\dots,i-1$. Clearly, the $2^v$ sets are disjoint and collectively contain all TB paths. Next, we introduce the concept of \emph{irreducible error event} (IEE) of $\TBP(i)$, the atomic TB path starting at state $i$ but not passing states $0,1,\dots, i-1$ in between. Thus, each path in $\TBP(i)$ can be reconstructed by concatenating the corresponding IEEs and then circular shifting the resultant path. Since the set of IEEs can be reused for equal or smaller trellis lengths, our collection algorithm will be efficient compared to the trivial algorithm. 

The paper is organized as follows. Sec. \ref{sec: the TBCCs} reviews the preliminaries of the TBCC and TB trellises, and Lou \emph{et al.}'s CRC design for ZTCCs. Sec. \ref{sec: CRC design for TBCC} introduces the partition of a TB trellis, IEEs, our DSO CRC design algorithm for TBCCs under low UERs, and a design example. Sec. \ref{sec: conclusion} concludes the entire paper.

\section{Preliminaries}
\label{sec: the TBCCs}

\subsection{Construction of the TBCC}
We briefly follow \cite{Ma1986} in describing a TBCC. For ease of understanding, consider a feedforward, $(n,1,v)$  convolutional code of rate $1/n$ and $v$ memory elements, albeit the design approach in this paper can be generalized to any feedforward, $(n,k,v)$ TBCC. For a binary information sequence of length $K$, $K\ge v$, we first use the last $v$ bits to initialize the convolutional encoder and ignore the outputs. Then the entire $K$-bit information sequence is fed into the encoder and the resultant $nK$-bit output is a TB codeword. As can be seen, the initial and final state of the codeword will be the same.  In this way the rate loss caused by termination in a ZTCC is eliminated. 
% In \cite{Ma1986}, the authors showed that any rate-$k/n$ ($k=1$ or $n-1$) TBCC is also a quasi-cyclic code with period $n$.

\subsection{Tail-Biting Trellises}

We follow \cite{Koetter2002} in describing the tail-biting trellises. Let $V$ be a set of vertices (or states), $\A$ the set of output alphabet, and $E$ the set of ordered triples or edges $(v,a,v')$, with $v,v'\in V$ and $a\in\A$.  In words, $(v,a,v')\in E$ denotes an edge that starts at $v$, ends at $v'$ and has output $a$.

% \begin{definition}[Conventional trellises, \cite{Koetter2002}]
% A conventional trellis $T=(V,E,\A)$ of depth $N$ is an edge-labeled directed graph with the following property. The vertex set $V$ can be partitioned as
% \begin{align}
% V=V_0\cup V_1\cup \cdots \cup V_N \label{eq: partition}
% \end{align}
% such that every edge in $T$ begins at a vertex of $V_i$ and ends at a vertex of $V_{i+1}$, $i=0,1,\dots,N-1$. The sets $V_0,V_1,\dots, V_N$ are called the vertex classes of $T$. The ordered index set $\I=\{0,1,\dots,N\}$ induced by the partition in \eqref{eq: partition} is called the time axis for $T$.
% \end{definition}

\begin{definition}[Tail-biting trellises, \cite{Koetter2002}]
A tail-biting (TB) trellis $T=(V,E,\A)$ of depth $N$ is an edge-labeled directed graph with the following property. The vertex set $V$ can be partitioned into $N$ vertex classes
\begin{align}
V=V_0\cup V_1\cup \cdots \cup V_{N-1} \label{eq: partition}
\end{align}
such that every edge in $T$ either begins at a vertex of $V_i$ and ends at a vertex of $V_{i+1}$, for some $i=0,1,\dots, N-2$, or begins at a vertex of $V_{N-1}$ and ends at a vertex of $V_0$.
\end{definition}

Geometrically, a TB trellis can be viewed as a cylinder of $N$ sections defined on some circular time axis. Alternatively, we can also define a TB trellis on a sequential time axis $\I=\{0,1,\dots,N\}$ with the restriction that $V_0=V_N$ so that we obtain a conventional trellis.

For a conventional trellis $T$ of depth $N$, a trellis section connecting time $i$ and $i+1$ is a subset $T_i\subseteq V_i\times \A \times V_{i+1}\subseteq E$ that specifies the allowed combination $(s_i, a_i, s_{i+1})$ of state $s_i\in V_i$, output symbol $a_i\in \A$, and state $s_{i+1}\in V_{i+1}$, $i=0,1,\dots,N-1$. Such allowed combinations are called trellis branches. A trellis path $(\bm{s},\bm{a})\in T$ is a state/output sequence pair, where $\bm{s}\in V_0\times V_1\times\cdots\times V_N$, $\bm{a}\in \A^N$. The code represented by trellis $T$ is the set of all output sequences $\bm{a}$ corresponding to all trellis paths $(\bm{s},\bm{a})$ in $T$. 

For a TB trellis $T$ of depth $N$, a TB path $(\bm{s},\bm{a})$ of length $N$ on $T$ is a \emph{closed} path through $N$ vertices. If $T$ is defined on a sequential time axis $\I=\{0,1,\dots,N\}$, then any TB path $(\bm{s},\bm{a})$ of length $N$ satisfies $s_0=s_N$.

In this paper, we only consider the TB trellis $T$ of depth $N$ satisfying $V_0=V_i$, $i=1,2,\dots,N-1$. Clearly, the TB trellis generated by the feedforward, $(n,1,v)$ convolutional encoder $\bm{g}(x)$ meets our condition.

% For a given minimal, non-catastrophic, rate-$1/n$ convolutional code $\bm{g}(x)=(g^{(1)}(x),\dots,g^{(n)}(x))$ with $v$ memory element, the tail-biting trellis defined on a sequential time axis $\I=\{0,1,\dots,N\}$ satisfies $T_i=T_0$ for all $i\in\I$. 

\subsection{System Model and Lou et al's CRC Design Method}

\begin{figure}[t]
\centering
\includegraphics[width=0.45\textwidth]{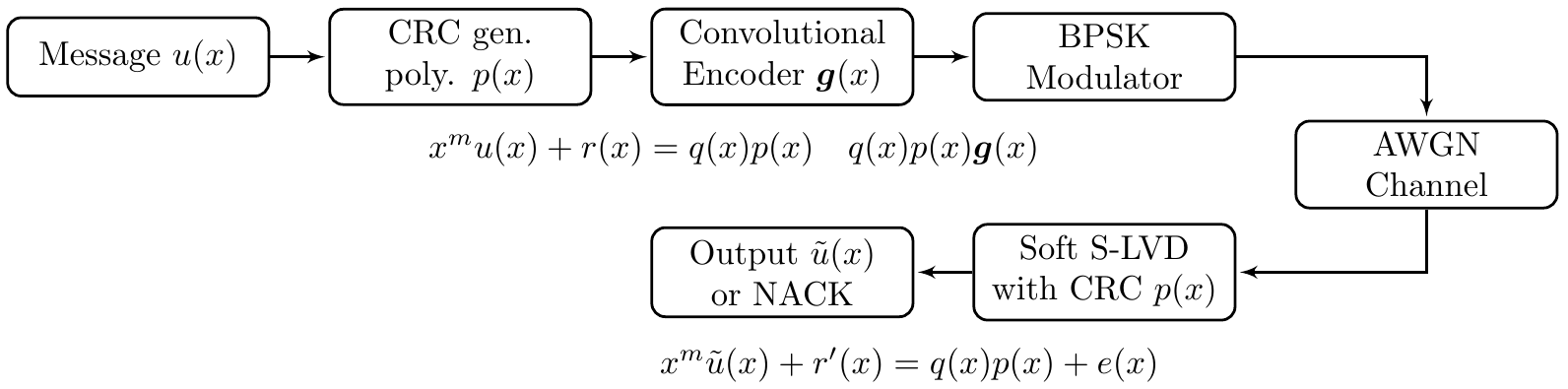}
\caption{Block diagram of a system employing CRC and convolutional codes.}
\label{fig: system model}
\end{figure}

We briefly follow \cite{Lou2015} in introducing their DSO CRC design scheme for a given ZTCC under a low target UER. 

The basic system model is depicted in Fig. \ref{fig: system model}. Let us consider a $K$-bit information sequence represented as a binary polynomial $u(x)$ of degree no greater than $K-1$. Then, the $m$ parity check bits are calculated as the remainder $r(x)$ of $x^mu(x)$ divided by a degree-$m$ CRC generator polynomial $p(x)$. Therefore, the $(K+m)$-bit sequence described by $x^mu(x)+r(x)$ is divisible by $p(x)$, i.e., there exists a unique polynomial $q(x)$ such that $x^mu(x)+r(x)=q(x)p(x)$. Hence, the CRC-coded sequence can be concisely expressed as $q(x)p(x)$. Let $\bm{g}(x)=(g^{(1)}(x),\dots,g^{(n)}(x))$ be the generator polynomial of a feedforward, rate-$1/n$ convolutional encoder.  After feeding the CRC-coded sequence $q(x)p(x)$ into the encoder, the output $q(x)p(x)\bm{g}(x)$ is the final codeword of the ZTCC. The transmitter sends the BPSK-modulated sequence of $q(x)p(x)\bm{g}(x)$ through an additive white Gaussian noise (AWGN) channel. After receiving the channel outputs, the serial list Viterbi decoder (S-LVD) produces the most likely message sequence $\tilde{u}(x)$ if a codeword passes the CRC check before reaching the maximum list size. Otherwise, a negative acknowledgement (NACK) is output. Performance analysis of S-LVD can be found in \cite{Yang2018}. An undetected error occurs if S-LVD erroneously identified a path corresponding to input sequence $q(x)p(x)+e(x)$, where $e(x)\ne0$ and is divisible by $p(x)$, as the maximum-likelihood (ML) path.  

The fundamental design challenge is to identify the optimal CRC for the ZTCC generated by $\bm{g}(x)$ such that the UER is minimized. Lou \emph{et al.} \cite{Lou2015} showed that if the target UER is low enough, the UER will be dominated by the smallest-distance undetected error. Therefore, designing the DSO CRC polynomial is equivalent to designing the CRC with the maximum undetectable minimum distance. This is the essential motivation of Lou \emph{et al}'s approach. In their design method, a CRC polynomial is removed from the candidate list if it possesses a smaller undetectable minimum distance or more undetected errors at the same distance. As we proceed to higher distances, the CRC candidate list is refined until only one candidate remains in the list. This candidate is the DSO CRC polynomial for the given ZTCC under the target UER.

% the problem is equivalent to designing a CRC polynomial $p(x)$ that maximizes the undetectable minimum distance of the overall concatenated code, i.e., the minimum distance at which an undetectable error event first occurs. Therefore, they proposed an algorithm which first collects a sufficient number of error events (or zero-terminated paths) of distances less than some threshold $\tilde{d}$ by performing a Viterbi search at all-zero state. Note that the smaller the distance is, the more likely the error event occurs. They exhaustively enumerate each degree-$m$ candidate CRC polynomial and count the number of error events each candidate detects in the order of increasing distances. In the end, the one that maximizes the undetectable minimum distance is the DSO CRC for ZTCCs.  

Note that the DSO CRC polynomial is always the one that minimizes the upper bound of the UER characterized by the full undetected distance spectrum. If the target UER is not low enough, the above CRC design procedure does not necessarily yield the DSO CRC polynomial.
% By considering more terms in the distance spectrum and their multiplicities, Lou's approach can also minimize UER in the more general case of noisier channels where minimizing UER is not necessarily equivalent to maximizing the undetectable minimum distance.

\section{Optimal CRC Design for the TBCC}
\label{sec: CRC design for TBCC}

In this paper, we consider the same system model as in Fig. \ref{fig: system model} except replacing ZTCCs with TBCCs. The primary distinction between the two types of convolutional codes is that a TB error event can start at a nonzero state and remain in nonzero states on the trellis. 

The fundamental DSO CRC design principle for a given convolutional code under a low target UER is analogous to that of Lou \emph{et al.}, which is to maximize the minimum distance at which an undetectable TB error event first occurs, (or undetectable minimum distance). Formally speaking, the degree-$m$ DSO CRC design procedure involves two steps. First, the \emph{collection algorithm} gathers a \emph{sufficient} number of error events of distances less than some threshold $\tilde{d}$ and stores them for future use. By ``sufficient'', we mean that the number of error events is enough to sieve the unique, degree-$m$ CRC  polynomial out of $2^{m-1}$ candidates\footnote{A CRC generator polynomial must have $1$ as coefficients for both the scalar term and the degree-$m$ term.}. Next, the \emph{search algorithm} initializes a list of $2^{m-1}$ CRC candidates. Iterating from distance $1$ to $\tilde{d}$, a candidate is removed from consideration if it possesses a smaller undetected minimum distance or more undetected errors at the same distance. Eventually, the last one in the list is the DSO CRC polynomial.
% Next, the \emph{search algorithm} enumerates $2^{m-1}$ CRC candidates. For each candidate, the search algorithm passes the error events in the order of increasing distance to the candidate CRC to perform a divisibility test and counts the total number of undetectable error events. In the end, the optimal CRC polynomial is the one that maximizes the undetectable minimum distance.

For TBCCs, the trivial collection algorithm is to perform Viterbi search at each initial state and then aggregate error events according to increasing distances. However, such an algorithm will be inefficient in designing DSO CRCs for a family of objective trellis lengths. The TB paths of one trellis length found by the trivial collection algorithm cannot be easily adapted to another trellis length.

% For TBCCs, the trivial collection algorithm is to perform Viterbi search at each initial state and then aggregate error events according to increasing distances. However, such an algorithm will be inefficient in designing high-degree CRCs since the list of error events includes tremendously many error events that could have been easily constructed by cyclic shifting and concatenating. Moreover, if the trellis depth changes, it becomes cumbersome to adapt the error events found by the trivial collection algorithm into other cases.

To enable the design for a family of objective trellis lengths, we propose an efficient collection algorithm that finds sufficient number of IEEs. These IEEs can be reused to reconstruct TB paths of any objective length via concatenation and circularly shifting the resultant path. The motivation of our collection algorithm originates from the following partitioning of TB trellises.

\subsection{Partitioning of the Tail-Biting Trellis}
\label{subsec: partitioning of trellises}
For a given feedforward, $(n,1,v)$ convolutional encoder $\bm{g}(x)$, let us consider the corresponding TB trellis $T=(V,E,\A)$ defined on a given sequential time axis $\I=\{0,1,\dots,N\}$. Since $T$ can also be represented by the union of TB paths (each corresponding to a TBCC codeword), we categorize each TB path according to the states through which it traverses. Formally speaking, let
\begin{align}
V^{(\pi)}_0=(\sigma_0,\sigma_1,\dots,\sigma_{2^v-1})
\end{align}
be a predetermined permutation of $V_0=\{0,1,\dots,2^v-1\}$. Define the set of TB paths w.r.t. $V_0^{(\pi)}$ as
\begin{align}
\TBP(\sigma_i)\triangleq &\big\{(\bm{s},\bm{a})\in V_0^{N+1}\times \A^N: s_0=s_N;  \notag\\
	\exists j\in\I& \text{ s.t. }  s_j=\sigma_i;\ \forall j\in\I,\ s_j\notin\{\sigma_0,\sigma_1,\dots,\sigma_{i-1}\} \big\},\notag\\
\forall i=&0,1,\dots,2^v-1.
\end{align}
In words, the set of $\TBP(\sigma_0)$ only contains TB paths that traverse through state $\sigma_0$; the set of $\TBP(\sigma_1)$ contains TB paths that traverse through state $\sigma_1$ but not $\sigma_0$; so on and so forth. Clearly, all sets $\TBP(\sigma)$, $\sigma\in V_0^{(\pi)}$, form a partition of the TB trellis $T$, i.e.,
\begin{align}
&\TBP(\sigma_i)\cap \TBP(\sigma_j)=\varnothing,\quad \text{if } \sigma_i\ne \sigma_j\\
&\bigcup_{\sigma\in V^{(\pi)}_0}\TBP(\sigma)=T.
\end{align}
An important property of the above decomposition is that each set $\TBP(\sigma)$ is closed under cyclic shifts.

\begin{theorem}\label{theorem: closure under cyclic shift}
Any cyclic shift of a TB path $(\bm{s},\bm{a})\in\TBP(\sigma)$ is also  a TB path in $\TBP(\sigma)$.
\end{theorem}

\begin{IEEEproof}
Since circularly shifting a TB path $(\bm{s},\bm{a})$ on a TB trellis $T$ defined on a given sequential time axis $\I=\{0,1,\dots,N\}$ is equivalent to circularly shifting $\I$ around $T$ defined on a circular time axis, this preserves the sequence of states (or vertices) through which the TB path $(\bm{s},\bm{a})$ traverses. Hence, the statement in Theorem \ref{theorem: closure under cyclic shift} holds.
\end{IEEEproof}

Inspired by the concepts of basis and linear combination in a vector space, we can consider the set of IEEs starting at state $\sigma$ as a basis from which each TB path of length $N$ in $\TBP(\sigma)$ may be constructed. The next section shows that this is accomplished by concatenating the IEEs and then circularly shifting the resultant TB path.

\begin{definition}[Irreducible Error Events]
For a TB trellis $T$ on sequential time axis $\I=\{0,1,\dots,N\}$, the set of irreducible error events $(\bm{s},\bm{a})$ at state $\sigma$ w.r.t. $V_0^{(\pi)}=(\sigma_0,\sigma_1,\dots,\sigma_{2^v-1})$ is defined as
\begin{align}
\IEE(\sigma_i)\triangleq\bigcup_{j=1,2,\dots,N}\overline{\IEE}(\sigma_i,j),\ \forall i=&0,1,\dots,2^v-1,
\end{align}
where
\begin{align}
    \overline\IEE(\sigma_i,j)\triangleq&\{(\bm{s},\bm{a})\in V_0^{j+1}\times \A^{j}: s_0=s_j=\sigma_i;\notag\\
	s_{j'}&\notin\{\sigma_0,\sigma_1,\dots,\sigma_i\} \text{ for all }j', 0<j'<j\}.
\end{align}
\end{definition}

\begin{algorithm}[t]
\caption{The Collection Algorithm}
\label{algorithm: collection}
\algrenewcommand\algorithmicrequire{\textbf{Input:}}
\algrenewcommand\algorithmicensure{\textbf{Output:}}
\begin{algorithmic}[1]
\Require The TB trellis $T$, threshold $\tilde{d}$, permutation $V_0^{(\pi)}$
\Ensure The list of IEEs $\List_{\IEE}(\tilde{d})=\{(\bm{s},\bm{a},\bm{u})\}$
\State Initialize lists $\List_{\sigma}$ to be empty for all $\sigma\in V_0^{(\pi)}$;
\For{$i\gets0,1,\dots,|V_0^{(\pi)}|-1$}
\State Perform Viterbi search at $\sigma_i$ on $T$ to collect list $\List_{\sigma_i}(\tilde{d})$ of all IEEs of distances less than $\tilde{d}$;
\EndFor
\State \Return $\List_{\IEE}(\tilde{d})\gets\bigcup_{\sigma\in V_0^{(\pi)}}\List_{\sigma}(\tilde{d})$;
\end{algorithmic}
\end{algorithm}

\begin{algorithm}[t]
\caption{The Search Algorithm}
\label{algorithm: search}
\algrenewcommand\algorithmicrequire{\textbf{Input:}}
\algrenewcommand\algorithmicensure{\textbf{Output:}}
\begin{algorithmic}[1]
\Require The length $N$, degree $m$, list of IEEs $\List_{\IEE}(\tilde{d})$
\Ensure The optimal degree-$m$ CRC gen. poly. $p(x)$
\State Initialize the list $\List_{\CRC}$ of $2^{m-1}$ CRC candidates, the empty list $\List_{\TBP}(d)$ of TBPs, $d=0,1,\dots,\tilde{d}-1$;
\For{$d\gets1,\dots,\tilde{d}-1$}
\State Construct new TBPs $(\bm{s},\bm{a},\bm{u})$ from $\List_{\IEE}(\tilde{d})$ s.t. $w_H(\bm{a})=d$, $|\bm{s}|=N$, via concatenating or cyclic shifting;
\State $\List_{\TBP}(d)\gets\List_{\TBP}(d)\cup\{(\bm{s},\bm{a},\bm{u})\}$;
\EndFor
\State $\Candidate(1)\gets\List_{\CRC}$;
\For{$d\gets1,\dots,\tilde{d}-1$}
\For{$p_i(x)\in \Candidate(d)$}
\State Pass all $\bm{u}(x)\in\List_{\TBP}(d)$ to $p_i(x)$;
\State $C_i\gets$ the number of divisible $\bm{u}(x)$ of dist. $d$;
\EndFor
\State $C^*\gets\min_{i\in\Candidate(d)}C_i$
\State $\Candidate(d+1)\gets\{p_i(x)\in\Candidate(d): C_i=C^*\}$;
\If{$|\Candidate(d+1)|=1$}
\State \Return $\Candidate(d+1)$;
\EndIf
\EndFor
\end{algorithmic}
\end{algorithm}

\begin{theorem}\label{theorem: irreducible error events}
Every TB path $(\bm{s},\bm{a})\in\TBP(\sigma)$ can be constructed from the IEEs in $\IEE(\sigma)$ via concatenation and cyclic shifting operations.
\end{theorem}

\begin{IEEEproof}
Let us consider $T$ as a TB trellis defined on a sequential time axis $\I=\{0,1,\dots,N\}$. For any TB path $(\bm{s},\bm{a})\in\TBP(\sigma)$ of length $N$ on $T$, we can first circularly shift it to some other TB path $(\bm{s}^{(0)},\bm{a}^{(0)})\in\TBP(\sigma)$ on $T$ such that $s_0^{(0)}=s_N^{(0)}=\sigma$.

Now, we examine $\bm{s}^{(0)}$ over $\I$. If $\bm{s}^{(0)}$ is already an element of $\IEE(\sigma)$, then there is nothing to prove. Otherwise, there exists a time index $j$, $0<j<N$, such that $s_j=\sigma$. In this case, we break the TB path $(\bm{s}^{(0)},\bm{a}^{(0)})$ at time $j$ into two sub-paths $(\bm{s}^{(1)},\bm{a}^{(1)})$ and $(\bm{s}^{(2)},\bm{a}^{(2)})$, where
\begin{align*} 
\bm{s}^{(1)}=&(s_0,s_1,\dots,s_j),\ \bm{a}^{(1)}=(a_0,a_1,\dots,a_{j-1}),\\
\bm{s}^{(2)}=&(s_j,s_{j+1},\dots,s_N),\ \bm{a}^{(2)}=(a_{j},a_{j+1},\dots,a_{N-1}).
\end{align*}
Note that after segmentation of $(\bm{s}^{(0)},\bm{a}^{(0)})$, the resultant two sub-paths, $(\bm{s}^{(1)},\bm{a}^{(1)})$ and $(\bm{s}^{(2)},\bm{a}^{(2)})$, still meet the TB condition. Repeat the above procedures on $(\bm{s}^{(1)},\bm{a}^{(1)})$ and $(\bm{s}^{(2)},\bm{a}^{(2)})$. Since the length of a new sub-path is strictly decreasing after each segmentation, the boundary case is the atomic sub-path $(\bm{s},\bm{a})$ of some length $j^*$ satisfying $s_0=s_{j*}=\sigma$, $s_{j'}\ne\sigma$, $\forall j'\in(0,j^*)$ which is clearly an element of $\IEE(\sigma)$. Thus, we end up obtaining sub-paths that are all elements of $\IEE(\sigma)$. Concatenating them yields the circularly shifted version of TB path $(\bm{s}^{(0)},\bm{a}^{(0)})$.
\end{IEEEproof}

Theorem \ref{theorem: irreducible error events} indicates that collecting IEEs starting at every state $\sigma$ is enough to reconstruct all TB paths in set $\TBP(\sigma)$. This underlies the collection and search algorithm we are about to propose. Note that the collection of IEEs only relies on the distance threshold $\tilde{d}$ assuming sufficiently long search depth. Once we collect \emph{all} IEEs of distance less than $\tilde{d}$, these IEEs can be reused to reconstruct TB path of distance less than $\tilde{d}$ and of any objective length.

% In fact, we can also prove Theorem \ref{theorem: irreducible error events} in the perspective of the corresponding state diagram, by demonstrating that each cycle traversing through state $\sigma$ can be decomposed into one or several irreducible cycles traversing through state $\sigma$. 

\subsection{The CRC Design Algorithm for the TBCC}
For the TB trellis $T$ of a feedforward, $(n,1,v)$ convolutional encoder $\bm{g}(x)$, let $(\bm{s},\bm{a},\bm{u})$ denote the triple of states $\bm{s}$, outputs $\bm{a}$ and inputs $\bm{u}$, where the inputs $\bm{u}$ are uniquely determined by state transitions $s_i\to s_{i+1}$, $i=0,1,\dots,N-1$. Motivated by the partitioning of $T$ and IEEs in Sec. \ref{subsec: partitioning of trellises}, we propose the collection algorithm and search algorithm to design the degree-$m$ DSO CRC polynomial $p(x)$, as demonstrated in Algorithm \ref{algorithm: collection} and \ref{algorithm: search}, respectively. In the pseudo-code description, we use $\bm{u}$ and $\bm{u}(x)$ interchangeably to denote the sequence and the corresponding polynomial, respectively. 

To visualize the process of the collection algorithm, consider the state diagram of the convolutional code, where each cycle in the state diagram with a length equal to the trellis depth represents a TB path. For a given ordering of states $V_0^{(\pi)}=(\sigma_0,\sigma_1,\dots,\sigma_{2^v-1})$, once the algorithm finds all IEEs starting from $\sigma_0$, the state diagram is reduced by removing $\sigma_0$ and the incoming and outgoing edges associated with it. The algorithm then finds the IEEs starting at $\sigma_1$ on the reduced state diagram. Repeating the above procedure, the collection algorithm is able to find all sets of IEEs.

% \begin{figure}[t]
% \centering
% \includegraphics[width=0.45\textwidth]{Figures/state_diagram_corrected.pdf}
% \caption{The state diagram of the feedforward, $(2,1,3)$ convolutional encoder $\bm{g}=(13, 17)_8$. The transition from state $\sigma_1$ to state $\sigma_2$ is labeled by an input/output pair $(\bm{u},\bm{a})$, where $\bm{u}\in\{0,1\}$, $\bm{a}\in\{0,1\}^2$.}
% \label{fig: state diagram}
% \end{figure}

The search algorithm first reconstructs each length-$N$ TB path of bounded distance through concatenation of IEEs and cyclic shift, and then finds the DSO CRC polynomial. The reconstruction step can be accomplished via dynamic programming. Specifically, let $\List(w,l)$ be the list of TB paths of weight $w$ and of length $l$, $0\le w< \tilde{d}$, $1\le l\le N$. Thus, given a new IEE $(\bm{s},\bm{a},\bm{u})$ of weight $w_H(\bm{a})$ and of length $|\bm{s}|$ satisfying $w_H(\bm{a})\le w$ and $|\bm{s}|<l$,
\begin{align}
    \List(w,l) = \List(w,l)\cup \{\List(w{-}w_H(\bm{a}),l{-}|\bm{s}|)+(\bm{s},\bm{a}, \bm{u})\},
\end{align}
where $+$ denotes the element-wise concatenation. Eventually, the lists $\List(w, N)$, $w<\tilde{d}$ stores all length-$N$ TB paths of distance less than $\tilde{d}$.

\subsubsection{Space complexity of the search algorithm} 
\begin{figure}[t]
\centering
\includegraphics[width=0.4\textwidth]{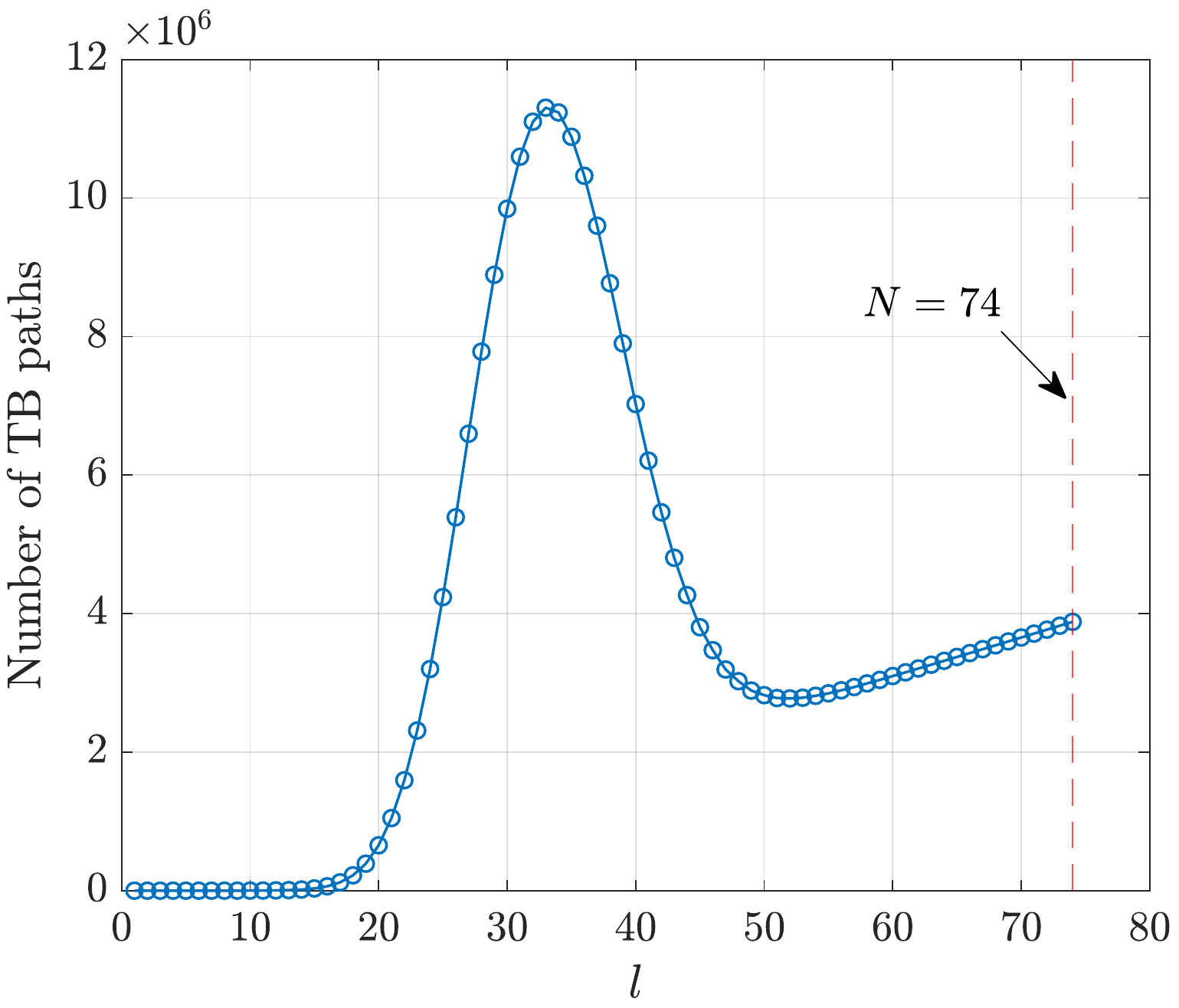}
\caption{The number of TB paths of length equal to $l$ vs. length $l$ for TBCC $(133,171)$ with $\tilde{d}=22$, $N=74$.}
\label{fig: growth of number of TB codewords}
\end{figure}

The space complexity is proportional to the total number of bits required to represent all TB paths in all lists $\List(w,l)$, $0\le w<\tilde{d}$, $1\le l\le N$. If the distance threshold $\tilde{d}$ is much less than the target length $N$, the growth of the number of TB paths of length equal to $l$ eventually becomes polynomial in $l$. Fig. \ref{fig: growth of number of TB codewords} shows the growth of number of TB paths of length $l$ for TBCC $(133,171)$ with $\tilde{d}=22$ and $N=74$. As can be seen, if $l\ge 3\tilde{d}$, the growth then becomes polynomial. This suggests that space complexity is polynomial in $N$ provided that $N>3\tilde{d}$.

\subsubsection{Choices of distance threshold $\tilde{d}$} In order to design the DSO CRC polynomial for a given TBCC, one has to select an appropriate $\tilde{d}$. Empirically, $\tilde{d}$ ranges from $2d_{\Free}$ to $3d_{\Free}$ for designing a CRC polynomial of degree $m\le10$.

\subsubsection{Choices of $V^{(\pi)}_0$} We note that in practice, the ordering of $V^{(\pi)}_0$ exerts a negligible influence on the space complexity. Hence, the natural ordering suffices for the DSO CRC design.

% Some remarks are given as follows. First, the collection requires the knowledge of permutation $V_0^{(\pi)}$. In practice, the permutation is selected based on the topology of the state diagram of the given convolutional encoder $\bm{g}(x)$. Second, in some special cases, we do not necessarily need to reconstruct all length-$N$ TB paths in the search algorithm. For example, if the CRC polynomial $p(x)$ is a minimal polynomial and inputs $\bm{u}(x)$ of a length-$N$ TB path is divisible by $p(x)$, so does its cyclic shift. Thus, we can directly calculate the number of distinct TB paths in this case.

\begin{table}[t]
\centering
\caption{Comparison of Undetected Distance Spectra between the Degree-$6$ DSO CRCs and the Suboptimal CRCs in \cite{Liang2019} TBCC $(13,17)$ and $N=70$ Bits. The DSO CRCs Are Highlighted. $A_d$ of Distances Between $8$ to $10$ Are All Zeros Thus Omitted. }
\setlength\tabcolsep{3.8pt}
\begin{tabular}{c|c|c|rrrrrrrr}
\hline
\multirow{2}{*}{$v$} & \multirow{2}{*}{TBCC}  & \multirow{2}{*}{CRC} & \multicolumn{8}{c}{Undetected Distance Spectra $A_d$} \\
\cline{4-11}
 	 & 	&   & $7$ & $11$ & $12$ & $13$ & $14$ & $15$ & $16$ & $17$  \\\hline
 \multirow{2}{*}{$3$} & \multirow{2}{*}{$(13,17)$} & 0x43 & $1$ & $8$ & $198$ & $758$ & $1114$ & $2814$ & $7375$ & $18473$ \\
     &  & \textbf{0x63} & $0$ & $0$ & $735$ & $0$ & $2310$ & $0$ & $13965$ & $0$\\\hline
\end{tabular}
\label{table: undetectable error dist spectra}
\end{table}

% \begin{table}[t]
% \centering
% \caption{Comparison of Undetected Distance Spectra between the Degree-$10$ DSO CRCs and the Suboptimal CRCs in \cite{Liang2019} for $3$ TBCCs and for $N=74$ Bits. The DSO CRCs Are Highlighted.}
% \setlength\tabcolsep{4.5pt}
% \begin{tabular}{c|c|c|rrrrrrr}
% \hline
% \multirow{2}{*}{$v$} & \multirow{2}{*}{TBCC}  & \multirow{2}{*}{CRC Poly.} & \multicolumn{7}{c}{Undetected Distance Spectra $A_d$} \\
% \cline{4-10}
%  	 & 	&   & $14$ & $15$ & $16$ & $17$ & $18$ & $19$ & $20$  \\\hline
%  \multirow{2}{*}{$6$} & \multirow{2}{*}{$(133,171)$} & 0x629 & $2$ & $0$ & $15$ & $0$ & $639$ & $0$ & $3400$ \\
%      &  & \textbf{0x475} & $0$ & $0$ & $10$ & $0$ & $898$ & $0$ & $2970$\\\hline
%  \multirow{2}{*}{$7$} & \multirow{2}{*}{$(247,371)$} & 0x61D & $6$ & $0$ & $3$ & $0$ & $172$ & $0$ & $1747$ \\
%  	& & \textbf{0x75D} & $0$ & $0$ & $0$ & $0$ & $247$ & $0$ & $1566$ \\\hline
%  \multirow{2}{*}{$8$} & \multirow{2}{*}{$(561,753)$} & 0x4CF & $0$ & $0$ & $1$ & $0$ & $204$ & $0$ & $775$\\
%   &   &  \textbf{0x579} & $0$ & $0$ & $0$ & $0$ & $17$ & $0$ & $1233$\\\hline
% \end{tabular}
% \label{table: undetectable error dist spectra}
% \end{table}

\begin{figure}[t]
\centering
\includegraphics[width=0.45\textwidth]{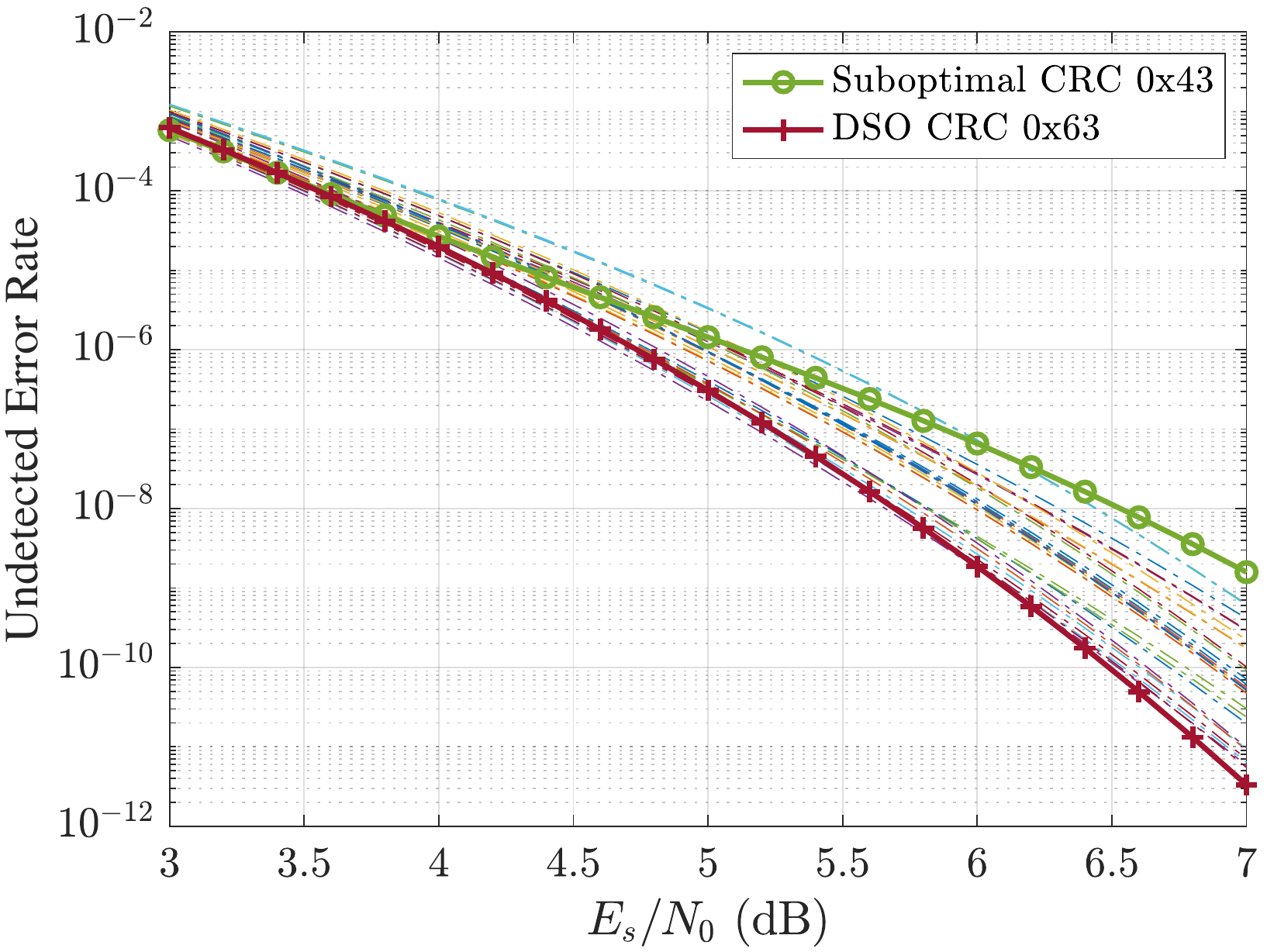}
\caption{Assume the target UER $P_e=10^{-10}$. The truncated union bound vs. SNR for all $32$ degree-$6$ CRC polynomial candidates for TBCC $(13,17)$ and $K=64$ bits. The degree-$6$ DSO CRC polynomial 0x63 and degree-$6$ suboptimal CRC polynomial 0x43 designed in \cite{Liang2019} are highlighted.}
\label{fig: truncated union bound}
\end{figure}

% \begin{figure}[t]
% \centering
% \includegraphics[width=0.45\textwidth]{Figures/Plot_ISIT_2020_TBCC_FER_VS_RCU.pdf}
% \caption{UER vs. Gap to RCU bound for the degree-$10$ DSO CRC polynomials and degree-$10$ suboptimal CRC polynomials designed in \cite{Liang2019}.}
% \label{fig: UER_vs_RCU}
% \end{figure}
\subsection{Example: Degree-$6$ DSO CRC for TBCC $(13,17)$}

As an example, we design the degree-$6$ DSO CRC polynomial for TBCC $(13,17)$ with $K=64$ bits under target UER $P_e=10^{-10}$. Since the UER is low enough, the UER in this regime will be dominated by the smallest undetected errors. 

Table \ref{table: undetectable error dist spectra} presents the undetected distance spectra up to $\tilde{d}=17$ for the degree-$6$ suboptimal CRC polynomial 0x43 designed in \cite{Liang2019} and our degree-$6$ DSO CRC polynomial 0x63 for TBCC $(13,17)$ and $K=64$ bits with overall  trellis length $N=K+m=70$. With the full undetected distance spectrum, the UER of a given CRC and TBCC can be upper bounded by the union bound of probability, namely,
\begin{align}
    P_e\le \sum_{d=1}^{d_{\max}}A_dQ\left(\sqrt{\frac{dE_s}{N_0}}\right), \label{eq: full upper bound}
\end{align}
where $d_{\max}$ is the maximum possible distance of the finite-length TBCC, and $Q(x)$ is the tail probability function of standard normal distribution. In practice, the full undetected distance spectrum can be computationally expensive. Instead, we will only calculate the bound in \eqref{eq: full upper bound} up to $\tilde{d}$ and such a bound is known as the \emph{truncated union bound}. Despite the resulting computational inaccuracy, the truncated union bound still serves as a good estimate in the low UER regime.

Fig. \ref{fig: truncated union bound} shows the truncated union bounds up to distance $\tilde{d}=17$ of all $32$ degree-$6$ candidate CRC polynomials for TBCC $(13,17)$. The curves corresponding to the suboptimal CRC 0x43 and the DSO CRC 0x63 are highlighted. As can be seen, the DSO CRC outperforms the suboptimal CRC by $2$ orders of magnitudes at $6.5$ dB, the SNR at which the DSO CRC attains the target UER of $10^{-10}$. However at $3$ dB, the DSO CRC 0x63 designed for $6.5$ dB performs worse than the CRC 0x43 and thus fails to remain optimal. This demonstrates that the DSO condition indeed depends on the operating SNR or target UER.

\section{Conclusion}
\label{sec: conclusion}

In this paper, we propose an efficient algorithm for designing DSO CRC polynomials for any specified TBCC for a low target UER. The algorithm is based on decomposing the TB trellis into several disjoint sets of TB paths that are closed under cyclic shifts. We also showed that the TB path in each set can be constructed from the IEEs via concatenation and cyclic shift. The use of IEEs enables the DSO CRC design for a family of trellis lengths (or the corresponding blocklengths). Our results demonstrate that for low target UER, DSO CRCs can significantly outperform suboptimal CRCs.

\bibliographystyle{IEEEtran}
\bibliography{IEEEabrv,references}

\begin{thebibliography}{10}
\providecommand{\url}[1]{#1}
\csname url@samestyle\endcsname
\providecommand{\newblock}{\relax}
\providecommand{\bibinfo}[2]{#2}
\providecommand{\BIBentrySTDinterwordspacing}{\spaceskip=0pt\relax}
\providecommand{\BIBentryALTinterwordstretchfactor}{4}
\providecommand{\BIBentryALTinterwordspacing}{\spaceskip=\fontdimen2\font plus
\BIBentryALTinterwordstretchfactor\fontdimen3\font minus
  \fontdimen4\font\relax}
\providecommand{\BIBforeignlanguage}[2]{{%
\expandafter\ifx\csname l@#1\endcsname\relax
\typeout{** WARNING: IEEEtran.bst: No hyphenation pattern has been}%
\typeout{** loaded for the language `#1'. Using the pattern for}%
\typeout{** the default language instead.}%
\else
\language=\csname l@#1\endcsname
\fi
#2}}
\providecommand{\BIBdecl}{\relax}
\BIBdecl

\bibitem{Ma1986}
H.~{Ma} and J.~{Wolf}, ``On tail biting convolutional codes,'' \emph{{IEEE}
  Trans. Commun.}, vol.~34, no.~2, pp. 104--111, February 1986.

\bibitem{Solomon1979}
G.~Solomon and H.~C.~A. Tilborg, ``A connection between block and convolutional
  codes,'' \emph{SIAM Journal on Applied Mathematics}, vol.~37, no.~2, pp.
  358--369, 1979.

\bibitem{Calderbank1999}
A.~R. {Calderbank}, G.~D. {Forney}, and A.~{Vardy}, ``Minimal tail-biting
  trellises: the golay code and more,'' \emph{{IEEE} Trans. Inf. Theory},
  vol.~45, no.~5, pp. 1435--1455, July 1999.

\bibitem{Koetter2002}
R.~{Koetter} and A.~{Vardy}, ``The structure of tail-biting trellises:
  minimality and basic principles,'' \emph{{IEEE} Trans. Inf. Theory}, vol.~49,
  no.~9, pp. 2081--2105, Sep. 2003.

\bibitem{Stahl1999}
P.~{Stahl}, J.~B. {Anderson}, and R.~{Johannesson}, ``Optimal and near-optimal
  encoders for short and moderate-length tail-biting trellises,'' \emph{{IEEE}
  Trans. Inf. Theory}, vol.~45, no.~7, pp. 2562--2571, Nov 1999.

\bibitem{Bocharova2002}
I.~E. {Bocharova}, R.~{Johannesson}, B.~D. {Kudryashov}, and P.~{Stahl},
  ``Tailbiting codes: bounds and search results,'' \emph{{IEEE} Trans. Inf.
  Theory}, vol.~48, no.~1, pp. 137--148, Jan 2002.

\bibitem{Wang1989}
Q.~{Wang} and V.~K. {Bhargava}, ``An efficient maximum likelihood decoding
  algorithm for generalized tail biting convolutional codes including
  quasicyclic codes,'' \emph{{IEEE} Trans. Commun.}, vol.~37, no.~8, pp.
  875--879, Aug 1989.

\bibitem{Cox1994}
R.~V. {Cox} and C.~E.~W. {Sundberg}, ``An efficient adaptive circular viterbi
  algorithm for decoding generalized tailbiting convolutional codes,''
  \emph{{IEEE} Trans. Veh. Technol.}, vol.~43, no.~1, pp. 57--68, Feb 1994.

\bibitem{Anderson1998}
J.~B. {Anderson} and S.~M. {Hladik}, ``Tailbiting map decoders,'' \emph{IEEE
  Journal on Selected Areas in Communications}, vol.~16, no.~2, pp. 297--302,
  Feb 1998.

\bibitem{Shao2003}
R.~Y. {Shao}, {Shu Lin}, and M.~P.~C. {Fossorier}, ``Two decoding algorithms
  for tailbiting codes,'' \emph{{IEEE} Trans. Commun.}, vol.~51, no.~10, pp.
  1658--1665, Oct 2003.

\bibitem{Chen2008}
{Tsao-Tsen Chen} and {Shiau-He Tsai}, ``Reduced-complexity wrap-around viterbi
  algorithm for decoding tail-biting convolutional codes,'' in \emph{2008 14th
  European Wireless Conf.}, June 2008, pp. 1--6.

\bibitem{Williamson2014}
A.~R. {Williamson}, M.~J. {Marshall}, and R.~D. {Wesel}, ``Reliability-output
  decoding of tail-biting convolutional codes,'' \emph{{IEEE} Trans. Commun.},
  vol.~62, no.~6, pp. 1768--1778, June 2014.

\bibitem{Niu2012}
K.~{Niu} and K.~{Chen}, ``{CRC}-aided decoding of polar codes,'' \emph{IEEE
  Commun. Lett.}, vol.~16, no.~10, pp. 1668--1671, October 2012.

\bibitem{Yang2018}
H.~{Yang}, S.~V.~S. {Ranganathan}, and R.~D. {Wesel}, ``Serial list viterbi
  decoding with {CRC}: Managing errors, erasures, and complexity,'' in
  \emph{2018 IEEE Global Commun. Conf. (GLOBECOM)}, Dec 2018, pp. 1--6.

\bibitem{Coskun2019}
M.~C. Co{\c s}kun, G.~Durisi, T.~Jerkovits, G.~Liva, W.~Ryan, B.~Stein, and
  F.~Steiner, ``Efficient error-correcting codes in the short blocklength
  regime,'' \emph{Physical Communication}, vol.~34, pp. 66 -- 79, 2019.

\bibitem{Liang2019}
E.~Liang, H.~Yang, D.~Divsalar, and R.~D. Wesel, ``List-decoded tail-biting
  convolutional codes with distance-spectrum optimal {CRCs} for {5G},'' in
  \emph{2019 IEEE Global Commun. Conf. (GLOBECOM)}, Dec 2019.

\bibitem{Lou2015}
C.~{Lou}, B.~{Daneshrad}, and R.~D. {Wesel}, ``Convolutional-code-specific
  {CRC} code design,'' \emph{{IEEE} Trans. Commun.}, vol.~63, no.~10, pp.
  3459--3470, Oct 2015.

\bibitem{Polyanskiy2010}
Y.~{Polyanskiy}, H.~V. {Poor}, and S.~{Verdu}, ``Channel coding rate in the
  finite blocklength regime,'' \emph{{IEEE} Trans. Inf. Theory}, vol.~56,
  no.~5, pp. 2307--2359, May 2010.

\end{thebibliography}

\end{document}